\documentclass{amsart}
\usepackage{amssymb}
\usepackage{verbatim}
\usepackage{mathrsfs}
\usepackage{dsfont}

\newtheorem{theorem}{Theorem}[section]

\newtheorem{corollary}[theorem]{Corollary}
\newtheorem{proposition}[theorem]{Proposition}

\theoremstyle{definition}
\newtheorem{definition}[theorem]{Definition}

\theoremstyle{remark}

\newcommand{\cA}{{\mathcal A}}

\newcommand{\cD}{{\mathcal D}}

\newcommand{\cF}{{\mathcal F}}
\newcommand{\cH}{{\mathcal H}}

\newcommand{\cM}{{\mathcal M}}
\newcommand{\qM}{{\mathfrak{M}}}
\newcommand{\Mb}{{\mathbb M}}

\newcommand{\cP}{{\mathcal P}}

\newcommand{\cO}{{\mathcal O}}

\newcommand{\cZ}{{\mathcal Z}}

\newcommand{\bC}{{\mathbb{C}}}

\newcommand{\Rn}{{\rm I\!R}} 
\newcommand{\Nn}{{\rm I\!N}} 
\newcommand{\Cn}{{\setbox0=\hbox{
$\displaystyle\rm C$}\hbox{\hbox
to0pt{\kern0.6\wd0\vrule height0.9\ht0\hss}\box0}}} 

\numberwithin{equation}{section}

\newcommand{\Tr}{\mathrm{Tr}}

\newcommand{\jed}{{\mathbb{I}}}
\begin{document}

\title{On quantum statistical mechanics; A study guide.}

\author{W. A. Majewski}

\address{Institute of Theoretical Physics and Astrophysics, The Gdansk University, Wita Stwosza 57,\\
Gdansk, 80-952, Poland and Unit for BMI, North-West-University, Potchefstroom, South Africa}
\email{fizwam@univ.gda.pl}

\date{\today}


\begin{abstract}
These notes are intended as an introduction to a study of applications of noncommutative calculus to quantum statistical Physics. Centered on noncommutative calculus we describe the physical concepts and mathematical structures appearing in the analysis of large quantum systems, and their consequences. These include the emergence of algebraic approach and the necessity of employment of infinite dimensional structures. As an illustration, a quantization of stochastic processes, new formalism for statistical mechanics, quantum field theory and quantum correlations are discussed.
\end{abstract}

\maketitle
\vspace*{1.5cm}
\noindent
\section{Basic ideas}
In these notes we will try and give an overview and road map to the area of quantum statistical mechanics without becoming too diverted by details. 
In contrast, we put a strong emphasis on evolution of calculus which is used in the description of statistical mechanics.
For much of the background we refer to the books of Omnes \cite{Omnes}, \cite{Omnes1} and Thompson \cite{thompson}, and for the more advanced material we refer to the books of Ruelle \cite{ruelle}, Emch \cite{emch}, Haag \cite{haag}, Takesaki \cite{Tak}, Terp \cite{terp} as well as for Bratteli and Robinson \cite{BR}.

To make our exposition abundantly clear we begin with a historical remark. Newton has given his principles for classical mechanics at the end of 17th century. However,  classical mechanics  blossomed into a rich mathematical theory only in the second half of 19th century. After a moment of reverie, we realize that although Newton and Leibniz introduced the basic principles of (classical) calculus it was Cauchy (around 1830';  albeit the ``epsilon-delta definition of limit'' was first given by Bolzano in 1817) who finally clarified the concept of limit and then Riemann (around 1860') who clarified the concept of integral. Consequently, in the second half of the 19th century the  principles of (classical) calculus were fully established. This gave the opportunity to transform  classical mechanics into a well-developed theory (Lagrange, Hamilton, Liouville...).
So with a mature theory of calculus available it took a few more decades to obtain a fully-fledged theory of classical mechanics. Subsequently, (classical) statistical mechanics has appeared as a combined development of classical mechanics and probability theory.

We will argue that a very similar situation has transpired in the 20th century, BUT now within the framework of quantum theory. The starting point was Heisenberg's equation of motion in Quantum Theory. He for the first time wrote a noncommutative derivation -- a commutator. (We remind that a derivation is a unary function satisfying the Leibniz product law.)
To see this, it is enough to note that a commutator satisfies the Leibniz rule!
This can be considered as an analogy of Newton's introduction of (classical) differentiation to write the equations of motion for a classical system. Then Heisenberg, Born, Jordan and Dirac realized that noncommutativity is the raison d'\^etre of quantum mechanics and they have introduced the so-called canonical quantization. It means that the basic relations of classical mechanics
\begin{equation}
\{p_i,q_j\} \propto \delta_{ij} 1, \qquad i,j = 1,2,3,...
\end{equation}
should be replaced by
\begin{equation}
\label{2}
[\hat{p_i}, \hat{q_j}] \propto \delta_{ij} \jed, \qquad i,j=1,2,3...
\end{equation}
where $\{\cdot, \cdot \}$ stands for the Poisson bracket while $[a,b] = ab-ba$ denotes the commutator.

BUT, the quantization procedure begs two serious questions:
\begin{enumerate}
\item In which terms can the relations (\ref{2}) be represented?
\item What can be said about uniqueness of the chosen representation?
\end{enumerate}

A brief answer to the first question says that the relations (\ref{2}) have no finite dimensional realization. Moreover, apart from Weyl's geometrical  quantization, (\ref{2}) are represented in terms of unbounded self adjoint operators acting on an infinite dimensional separable Hilbert space. We emphasize that in Weyl's quantization, using the functional calculus, one considers the unitary operators $V(t) = e^{i\hat{p}t}$ and $U(s) = e^{i\hat{q}s}$.

But a rigorous study of the Schr{\"o}dinger representation of canonical commutation relations for a finite degree of freedom leads to $^*$-algebra of unbounded operators, see Example 2 in \cite{power}. Moreover, Wightman's formulation of quantum field theory and the theory of Lie algebras lead to the scheme for a description of a physical system which is based on unbounded operators. Although mathematical aspects of algebras of unbounded operators have been analyzed in many details, see \cite{AIT}, \cite{Schmud}, \cite{Bag}, it is well known that formal calculations can be misleading , see Section VIII.5 in \cite{RSI}.

 Generally, it would seem that in Quantum Mechanics one can distinguished two schemes for a description of a physical system, cf. \cite{Bor1}. The first one, just described, uses unbounded operators. The second one uses bounded operators. The idea of introducing the norm topology on the set of observables was strongly advocated by I. Segal \cite{Segal}. To argue in favor of this idea one can say that in a laboratory a physicist deals with bounded functions of observables only! However, as it was already remarked by Borchers \cite{Bor1} in this method \textit{``some detailed information about a physical system is usually lost''}. Furthermore, this scheme admits ``non-physical states'' having badly defined entropy, see \cite{[3]} and references given there.

Here, we will argue that non-commutative integration theory offers the third scheme lying between the above discussed approaches. Besides other technical conditions it relies of selecting ``more'' regular unbounded operators, where ``more'' regular means $\tau$-measurability (see next pages for definitions and details).
Consequently, as it will be described, one is getting a very well behaved $^*$-algebra of unbounded operators. Moreover, bounded functions of self-adjoint elements of this algebra are elements of certain algebra of bounded operators.

Turning to the second question we should recall the so-called uniqueness theorem, attributed to von Neumann, Weyl, and Rellich. This theorem says that the answer to the second question takes into account the nature of the considered system. More precisely, a system will be called small if it has finite number of degrees of freedom. On the contrary, a system with an infinite number of degrees of freedom is called a large system. 

The uniqueness theorem states that for small systems, the relations (\ref{2}), up to unitary equivalence, have a unique representation. It is worth pointing out that this property is the keystone in formulation of the Dirac's formalism of quantum mechanics. We remind that the basis of that formalism is the pair 
\begin{equation}
\label{3}
\left( B(\cH), {\cF}_T(\cH) \right)
\end{equation}
 where $B(\cH)$ denotes all bounded linear operators on a separable, infinite dimensional Hilbert space $\cH$. $\cF_T(\cH)$ stands for trace class operators on a Hilbert space $\cH$. In particular, density matrices describing (quantum) states form a convex (generating) subset of ${\cF}_T(\cH)$.

For large systems the situation is very different. There are \textit{plenty} of non-equivalent representations of the relations (\ref{2}) when the number of degrees of freedom is infinite. The crucial point to note here is that both statistical mechanics as well as field theory are par excellence theories of large systems!

This fact was recognized in the fifties of the 20th century with the discovery of the so called ``strange representations''. Further, it was observed that carrying out the quantization of large systems on the basis of Dirac's formalism can lead to serious difficulties. To give illustrative examples, we firstly mention
problems associated with the Fock representation. The Fock representation was introduced in 1932 and subsequently fully elaborated by Cook in 1953. It is probably the best known scheme for a description of infinite quantum systems. But within this representation, one is able to describe only quasi-free systems. In other words, we cannot describe interacting particles. Furthermore, as was shown by van Hove in the fifties, (\cite{vH1}, \cite{vH2}, see also subsection $Ie$ in \cite{emch}), there does not exist a non-trivial perturbation calculus within the Fock representation.
We can not resist mentioning that a perturbation calculus is the main tool for calculations in Dirac's formalism. Finally, it is worth pointing out that the interaction picture does not exist in an interacting relativistic quantum field theory; this is the essence of the Haag theorem, see Section II.1.1 in \cite{haag} and/or Section 3.1.d in \cite{emch}.

Turning to the second example we wish to discuss the quantum Gibbs Ansatz. The Gibbs Ansatz was designed to describe a (classical) canonical equilibrium state and, up to normalized constant, is given 
by $e^{-\beta H}$. Here, $H$ stands for the Hamiltonian of the considered system, and $\beta$ is the ``inverse'' temperature. We emphasize that this is the basic ingredient of classical statistical physics. The quantization of $e^{-\beta H}$ means that now $H$ is the Hamilton operator and to have a quantum state, we require that $e^{-\beta H}$ should be a trace class operator.
But this is the case when, at least, necessary conditions are satisfied: $H$ has pure point spectrum with accumulation point at infinity. Unfortunately, even Hamiltonian of the Hydrogen atom does not fulfill this requirement!

Consequently, we arrived at the conclusion that in accordance with the second part of the (non)uniqueness theorem, one should take as a starting point, algebraic structures which are different from $B(\cH)$.

Before proceeding further let us pause to describe briefly possible algebras (other than $B(\cH)$) which could be useful for a description of a quantum large system.

We start with the notion of $^*$-Banach algebra. It is a Banach space $\mathfrak{B}$ equipped with multiplication and involution. Both operations are continuous with respect to the topology induced by the norm. If $a\cdot b = b\cdot a$ for $a,b \in \mathfrak{B}$ then $\mathfrak{B}$ is called commutative.
When, the norm satisfies the following extra condition: $\|a^*a\| = \|a\|^2$ then such a $^*$-Banach algebra is called a $C^*$-algebra and will be denoted by $\mathfrak{A}$. A von Neumann algebra $\mathfrak{M}$ is a concrete $C^*$-algebra $\mathfrak{A}$ (so $\mathfrak{A} \subset B(\cH)$ for a Hilbert space $\cH$) which is closed with respect to the weak operator topology.
The important point to note here is that every commutative von Neumann algebra is isomorphic to $L^{\infty}(X)$ for some measure space $(X, \mu)$ and conversely, for every $\sigma$-finite measure space $X$, the *-algebra $L^{\infty}(X)$ is a von Neumann algebra. Here, $L^{\infty}(X)$ stands for all (essentially) bounded functions on $X$.
Consequently, noncommutative von Neumann algebras provide nice starting point for the theory of noncommutative integration.
We complete this brief list of algebraic structures with the definition of $O^*$-algebra. It will be used in description of Wightman's postulates. $O^*$-algebra is a $^*$-algebra $\cA$ of linear operators defined on a common dense subspace $\cD$ of a Hilbert space $\cH$ and leaving $\cD$ invariant. The multiplication in $\cA$ is composition of operators while the involution $a \mapsto a^{\dagger}$ in $\cA$ i defined by $a^{\dagger} = a^*|_{\cD}$
where $a^*$ is the usual Hilbert space adjoint.

In the thirties of the last century, von Neumann and  Murray gave a classification of von Neumann algebras. To describe this classification, we first of all, recall the definition of the center $\cZ(\mathfrak{M})$ of the algebra $\mathfrak{M}$:
$$\cZ(\mathfrak{M}) = \{a \in \mathfrak{M}: ab = ba \quad\rm{for \ all} \quad b\in \mathfrak{M} \}.$$
$\mathfrak{M}$ is called a factor if $\cZ(\mathfrak{M}) = \Cn \jed$. Von Neumann \cite{vN1949} showed that every von Neumann algebra on a separable Hilbert space is isomorphic to a direct integral of factors. This decomposition is essentially unique. Therefore, to give the aforesaid classification of von Neumann algebras, one can restrict oneself to factors.

 One can distinguish three types of factors. The first type, denoted by I, consists of algebras of all linear bounded operators on a Hilbert space $\cH$. If $\dim \cH = n < \infty$, then one gets $M_n(\bC)$ - the algebra of $n \times n$ matrices with complex entries. Such factors are denoted by $I_n$. When $\cH$ is a separable infinite dimensional Hilbert space, then
we get the basic ingredient of Dirac's formalism -- $B(\cH)$. These $B(\cH)$ algebras are equipped with the canonical trace $\Tr$, i.e. a partially defined positive, linear functional such that $\Tr ab = \Tr ba$ for any $a,b \in B(\cH)$. $\Tr$ is defined as the sum of the diagonal elements of a matrix representation of $a \in B(\cH)$.

The second type, denoted by II, roughly speaking, consists of algebras such that their projections are of a specific type; more precisely, there are no minimal projections, but there are non-zero finite projections. Type I and II are called semifinite.  Such algebras have the important common property that
they can be equipped with a trace. We emphasize that a given trace on a semifinite algebra can be different from the canonical one  which was described for algebras of type I.

Finally, there are also type III  factors. The important property of these factors is that they cannot be equipped with a non-trivial trace, see for instance Section 2.7.3 in \cite{BR}. For a deeper discussion we refer the reader to \cite{Tak}.

For a long time type III algebras were, especially in mathematical physics, considered as exotic ones. But, in 1967, this point of view was completely abandoned. In his work, Powers \cite{pow} was
studying representations of uniformly hyperfinite algebras. In very ``physical'' terms his results can be expressed as an analysis of a
one dimensional spin chain. Such a model consists of an infinite number of sites, with the algebra $M_2(\bC)$ associated to each site. Thus, local observables associated with a site are given by elements from $M_2(\bC)$. Local equilibrium at each site is given by a $2 \times 2$ matrix of the form $Z^{-1} e^{- \beta H_{loc}}$ where $Z$ is the normalizing constant, and $H_{loc} \in M_2(\bC)$ is the local Hamiltonian associated with a site. Studying the thermodynamical limit of the above system, Powers has shown that for $\beta \notin \{0, \infty \}$ the equilibrium representations lead to type III of von Neumann algebras. Moreover, if $\beta \neq \beta^{\prime}$ one gets non-equivalent type III factors. Consequently, he has shown that von Neumann algebras of this type form a large family and that they can be labeled by a ``physical'' parameter.

The subsequent results obtained by Araki-Woods, Hugenholtz et al, and others, have shown that this type of von Neumann algebra is typical in the study of large systems, see \cite{Yngvason}. We emphasize that this is in perfect harmony with the second part of the (non)-uniqueness theorem; quantization of large systems leads to different algebras than $B(\cH)$!

 \vskip 1cm

As it was mentioned at the beginning, the precise description of limit and integral in classical calculus was steering the development of classical mechanics as well as statistical mechanics.
Here we wish to describe the analogous process, but now for the quantum theory.

In late thirties of the last century, von Neumann realized that non-commutative integration should play an essential role in quantum theory. To start with he proposed to carry out noncommutative integration by using tricky
norms defined on matrix algebras, see \cite{vNeu}. 

But, the essential step was independently done by I. Segal \cite{seg} and J. Dixmier \cite{dix} in the early fifties. They generalized the concept of integration to much more general algebras. For semifinite von Neumann algebras the theory of noncommutative integration was completed by E. Nelson in 1974, see \cite{nel}. It is very important to note that as a first step it was necessary to define the concept of noncommutative measurable operators (quantum counterpart of measurable functions). To this end the concept of trace is necessary. Consequently, the theory of noncommutative integrals was done for von Neumann algebras of type I and II.

We can not resist mentioning one striking feature of that theory. Restricting to $B(\cH)$, one can show that all (noncommutative) measurable operators are bounded!  This is not true for other algebras. Thus, this strange result indicates how Dirac's formalism is exceptional. In other words, the noncommutative calculus for small system differs very much from that which is applicable to large systems.

To proceed with an analysis of large systems, the aforesaid theory of noncommutative integration should be generalized to type III von Neumann algebras. This was achieved, firstly by  Haagerup's seminal paper (1977), and secondly by contributions given mainly by Takesaki, Connes, Hilsum, Araki-Masuda, Kosaki, and Dodds, Dodds, de Pagter. The best general reference here is \cite{terp}. For a deeper discussion we refer the reader to \cite{Tak2} and \cite{Connes}.
 The essential step of the above generalization relies on the construction of a (much) larger von Neumann algebra, the so called crossed product $\mathfrak{M} \rtimes_{\sigma} \Rn$ where $\mathfrak{M}$ is the original algebra and $\sigma$ stands for the modular action. The important point to note here is that $\mathfrak{M}$ is a proper subset of $\mathfrak{M} \rtimes_{\sigma} \Rn$. Moreover, $\mathfrak{M}$ can be easily identified in $\mathfrak{M} \rtimes_{\sigma} \Rn$ as the family of fixed points of certain canonical map. 

The principal significance of the larger algebra $\mathfrak{M} \rtimes_{\sigma} \Rn$ is that this crossed product is a semifinite super algebra for the type III algebra $\mathfrak{M}$. Hence, $\mathfrak{M} \rtimes_{\sigma} \Rn$ can be equipped with a (semifinite) trace. Consequently, one can define noncommutative measurable operators. Then by a concrete selection of certain subsets of noncommutative measurable operators one arrives at noncommutative counterparts of (classical function) spaces, e.g. noncommutative $L^p$ or Orlicz spaces.

To complete the above brief exposition on noncommutative calculus, we must add that a complete account on (noncommutative) derivations was given in \cite{Brat}. For an illustration of how derivations may be used for a study of quantum dynamical systems, we refer the reader to \cite{sakai}.
However, as we will not use these facts in any essential way, this topic will be dropped.

To sum up:

\vspace{0.5cm}

\textit{For (large) quantum systems, in the nineties of the 20th century, we got a situation which can be considered as analogous to the one which pertained, for classical physics, at the end of the 19th century. Thus, we are in position to employ the just presented noncommutative calculus for a description of large systems, i.e. to analyze quantum statistical mechanics as well as quantum field theory. This gives an opportunity to transform quantum statistical mechanics as well as quantum field theory into a well-developed theory. Finally, it should be now clear that the calculus used in Dirac's formalism is not well adapted for a study of large systems. Moreover, a genuine quantum system can not be described within finite dimensional structures. In particular, the proper description of quantum systems can not rely on factors $I_n$, $n < \infty$!}

\section{Applications}
Now we are in a position to indicate how the calculus described above may be applied to a study of large systems.

\subsection{A clarification of old problems.} 
The first important consequence of the aforesaid framework is that it yields a better understanding of difficulties which appeared, in the fifties of the 20th century, in the study of large quantum systems, (cf. the first section). In particular, we have seen that from the noncommutative integration point of view, the algebra $B(\cH)$
is a very special one. Moreover, the calculus based on the pair $\left(B(\cH), \cF(\cH) \right)$ is not well adapted for a study of large quantum systems. 
We note here that a linear positive functional conditioned only by its behaviour on bounded observables can exhibit ``unphysical'' properties, eg. it can lead to problems with a definition of entropy. A more detailed discussion of this problem will be postponed until subsections (\ref{2.3}) and (\ref{2.4}). 

Turning to ``strange representations'' we note that a non-trivial interaction can lead to a change of the Fock representation to another, non-equivalent one.
Consequently such facts as van Hove's observation on perturbation calculus carried out within Fock space framework and Haag's theorem for quantum field theory are not unexpected results! A clarification of these problems was done within the Haag-Kastler approach, for details see \cite{haag}. For a recent account of a locally covariant quantum field theory we refer the reader to \cite{BFV}. An application of non-commutative calculus to quantum fields will be given in subsection (\ref{2.4}).

Closing this subsection, we want to say that difficulties relating to the Gibbs Ansatz were solved by developing KMS theory; i.e.\ a general $C^*$-machinery was employed! For details see vol. II of Bratteli, Robinson's book \cite{BR}.
\subsection{ Quantization of Markov-Feller processes.}
Markov-Feller processes constitute an important subset in the family of (classical) stochastic processes (see \cite{ligget}). A hallmark of these processes is the one-to-one correspondence with Markov semigroups. In turn, a Markov semigroup is uniquely determined by its infinitesimal generator. But, an infinitesimal generator of a Markov semigroup associated with a Markov-Feller process  has an explicit form which is given in terms of (classical) $L^2$-spaces. To fully elaborate the description of infinitesimal generators corresponding to Markov-Feller processes, classical $L^p$-spaces and the interpolation strategy have proven to be very effective tools. As all ingredients of the aforesaid strategy have their noncommutative counterparts, the quantization of Markov-Feller processes has been a straightforward task, see \cite{MZ1} - \cite{MZ4}. 

Working within the Haagerup theory on noncomutative $L^p$-spaces, the following results were obtained:
\begin{enumerate}
\item Noncommutative $L^p$-spaces for quantum lattice models were described.
\item Quantum counterparts of classical infinitesimal generators of Markov semigroups associated to Markov-Feller processes were obtained and studied. It is worth pointing out that both jump-type and diffusion-type processes were analyzed.
\item Concrete illustrative models of quantum dynamical systems were given, cf \cite{MOZ}.
\end{enumerate}
But, one may ask whether such quantized dynamics exhibits a stability and/or one is able to describe ``return to equilibrium'' for such dynamical systems. It seems that the most important tools in such studies are log Sobolev inequalities, see A. Guionnet-B. Zegarlinski's thoroughgoing review \cite{GZ} for a recent account of that theory and a comprehensive bibliography. But, one may conjecture  that the theory of (noncommutative) $L^p$-spaces is not well adapted for such studies, see \cite{BG} and the next subsection.

\subsection{ Statistical Mechanics and Boltzmann theory.} \label{2.3}
As it was mentioned, in the standard approach, the basic mathematical ingredient of (quantum) statistical mechanics is the dual pair (\ref{3}) modeling the states and observables of the system under consideration. But the crucial points to note here are the following observations:
\begin{enumerate}
\item For any $a \in B(\cH)$ and any $\varrho \in {\cF}_T(\cH)$ one has that for any $n \in \Nn$,  
$Tr \varrho a^n < \infty$. Consequently, in the standard approach to statistical mechanics we are employing observables having all moments finite. Such observables are called regular. We emphasize that the same can be said for classical systems.
\item The pair (\ref{3}) allows bounded observables only. But in both cases (classical and quantum) typical observables are unbounded.
\end{enumerate}

The above observations suggest the necessity for a more general setting which allows unbounded observables and, at the same time, preserves the property of finiteness of all moments. Such a more general setting was proposed in \cite{[1]}, \cite{[2]}. To describe the above generalization we need some preliminaries.

The classical $L^p$-spaces form a subset of a broader class of Banach spaces - the class of Orlicz spaces. Orlicz spaces are defined by selecting a subset of measurable functions by means of an appropriate Young's function. In these notes we will need two concrete Orlicz spaces: $L^{\cosh - 1}$ and $L\log(L+1)$. They are defined by the corresponding Young's functions: $x \mapsto \cosh(x) -1$ and $x \mapsto x\log(x+1)$. In particular, $L^{\cosh - 1}$ is the subset of measurable functions $f$ such that $\int(\cosh(f) - 1) d\mu < \infty$, where $\mu$ is a measure fixed by a considered model. The principal significance of $L^{\cosh - 1}$-space stems from the Pistone-Sempi result \cite{PS}: \textit{classical regular observables are described by $L^{\cosh - 1}$-space.}
Furthermore, $L\log(L+1)$-space is an isomorphic copy of the dual space $L^{\cosh - 1}$. It is important to note that this dual space is defined by the entropic-type function $x \mapsto x\log(x+1)$. We have, \cite{[1]}- \cite{[2]},
\begin{proposition}  The dual pair $\left(L^{\cosh - 1}, L \log(L+1)\right)$ provides the basic mathematical ingredient for a description of a general (regular) classical system.
\end{proposition}
To support the above claim we note, (for details see \cite{[1]}, \cite{[2]}, \cite{[3]}):
\begin{enumerate}
\item The Pistone-Sempi result \cite{PS} says that $L^{\cosh - 1}$ is well adapted for a description of classical regular observables.
\item The entropy is much better defined for $f \in L\log(L+1)$. It is well known that the condition $f \in L^1$ is not sufficient to guarantee well definiteness of the entropy, see \cite{Bour}, Chapter IV, \S 6, Exercise 18.
\item In the modern theory of Boltzmann's equation, the space $L\log(L+1)$ appears as a condition for the existence of weak solutions of Boltzmann's  equation (for a large class of kernels), see \cite{Vil}.
It is worth pointing out that the condition $f \in L^1$ is too weak to guarantee the existence of solutions of Boltzmann's equation!
\item log-Sobolev inequalities, the basic tool in an analysis of the stability of dynamics, can be written as Poincar\'e-type inequalities on $L\log(L+1)$ \cite{BG}.
\end{enumerate}
Turning to quantum systems, the noncommutative integration theory allows one to (see also \cite{DDdP} and \cite{L}),
\begin{enumerate}
\item define quantum counterparts of $L^{\cosh - 1}$ and $L\log(L+1)$ (for simplicity, we will denote them by the same symbols),
\item to show that the entropy is also much better defined,
\item to study log-Sobolev inequalities in the quantum setting.
\end{enumerate}

Although the quantization of Boltzmann's equation is not clear, the approach to quantum statistical mechanics based on the quantum dual pair $\left( L^{\cosh - 1}, L\log(L+1) \right)$ offers the possibility to quantize a large class of classical dynamical maps as well as to lift dynamical maps defined on the algebra of bounded observables $\mathfrak{M}$ to well defined maps on $L^{\cosh - 1}$, \cite{[4]}. To this end, one is in the first case using the interpolation scheme based on the DDdP approach, cf. \cite{DDdP}. This follows from combining the Orlicz result that the Orlicz space $L^{\Phi}(0, \infty)$ ($\Phi$ stands for a Young's function) is an interpolation space in the couple$\left( L^1(0, \infty), L^{\infty}(0, \infty) \right)$, and noncommutative interpolation for such couples given in \cite{[12]}. To lift the dynamics $T_t : \mathfrak{M} \to \mathfrak{M}$, it was shown \cite{[4]} that assuming additionally a condition acceptable from a Physics point of view and working within Haagerup's approach to noncommutative integration, one is able to lift $T_t$ to a well defined map on $L^{\cosh - 1}(\mathfrak{M})$. In particular

\begin{theorem}
If $T$ is a completely positive map on $\mathfrak{M}$ satisfying the Detailed Balance Condition, then the extension $\widetilde{T}$ of $T$ on $\mathfrak{M} \rtimes_{\sigma} \Rn$ canonically induces an action on $L^{\cosh -1}(\mathfrak{M})$.
\end{theorem}

\subsection{Applications of Orlicz spaces to quantum fields} \label{2.4}
In this subsection we will look at the second type of large systems -- quantum fields. Our brief exposition for the description of these large systems will be done within the framework set by the connection of Wightman quantum field theory (see \cite{W&S}, \cite{jost}) with the theory of local nets of $C^*$ or von Neumann algebras (\cite{haag}, \cite{araki}, see also \cite{B&H}; for all details see \cite{LM17}).

In the description of the first type of large systems -- quantum statistical mechanics -- the finiteness of all moments was the crucial requirement, cf. the previous subsection. The important point to note here is a consequence of the first postulate of Wightman theory. We remind (cf. \cite{araki}) the first postulate of Wightman theory: there are operators $\phi_1(f), ...,\phi_n(f)$, where $f_i \in C^{\infty}_0(\Mb)$ is a $C^{\infty}$-function with compact support on the Minkowski space $\Mb$.
Each $\phi_j(f)$ and its hermitian conjugate operator $\phi^*_j(f)$ are defined, at least, on a common dense subset $\cD$ of the Hilbert space $\cH$. Moreover:
\begin{equation}
\phi_j(f) \cD\subseteq \cD
\end{equation}
and
\begin{equation}
\phi_j(f)^* \cD \subseteq \cD
\end{equation}
for any $f \in C^{\infty}_0(\Mb)$ and any $j = 1,...,n$. 
Another way of stating this postulate is to say that the set of field operators constitute a $O^*$-algebra. To begin with a selection of measurable field operators we note:
 for any $m \in \Nn$, and any $u,v \in \cD$ one has
\begin{equation}
(u, \phi^m(f)v) \in \Cn.
\end{equation}
 In other words, one can say that (unbounded) field operators have finite moments (on the common domain).

It is worth pointing out that this observation can be considered as an ``invitation'' to Orlicz space approach --  this feature of quantum statistical formalism was the starting point for developing the new approach to large systems, see \cite{[2]}, \cite{[3]}.

Hence one may ask whether the new formalism for quantum statistical mechanics can be extended to quantum field theory. To show that this is a case it is necessary to make more precise the requirement that a Wightman field is associate to a local net of von Neumann algebras. 

By a local net of von Neumann algebras we mean an assignment
$$\cO \mapsto \cA(\cO)$$
of regions $\cO$ in the Minkowski space $\Mb$ (or more generally in Lorentzian manifold $M$) to von Neumann algebras $\cA(\cO)$ on the Hilbert space $\cH$ of the field operators such that the usual conditions of isotony, locallity and covariance are fulfilled.

A field operator can be associated to a net in different ways. We shall use the following one.

Let $\cP$ be a family of operators with a common dense domain of definition $\cD$ in a Hilbert space $\cH$  such that if $\phi \in \cP$ then also $\phi^*|_{\cD} \equiv \phi^{\dagger} \in \cP$. The weak commutant $\cP^w$, of $\cP$ is defined as the set of all bounded operators $C$ on $\cH$ such that $(v,C\phi w) = (\phi^{\dagger}v, C w)$, for all $v,w \in \cD$.

For simplicity of our arguments we will restrict ourselves to one type of real scalar field $\phi$; i.e. $\phi(f)^*$ coincides with $\phi(\overline{f})$ on $\cD$. Furthermore, apart from the Wightman postulates we assume:

\begin{enumerate}
\item[{(A1)}]\label{I} $\cP(\cO^p_q)^w$ is an algebra for any double cone $\cO^p_q \equiv \{x; p-x \in V_+, x - q \in V_+ \}$, where $V_+ = \{ \rm{ positive \ timelike \ vectors} \ \}$.

\item[{(A2)}]\label{II} The vacuum vector $\Omega$ is cyclic for the union of $\cP(D^{\prime})^w$ over all double cones $D$, where $D^{\prime}$ is the causal complement of $D$.
\end{enumerate}

The following theorem is taken from \cite{araki} , but stems from results given in \cite{DSW}, \cite{buch1}.

\begin{theorem}
\label{2.3a}
Assume both conditions (A1) and (A2). For each double cone $D$, define
\begin{equation}
\qM(D) = \left( \cP(D)^w \right)^{\prime}.
\end{equation} 
Then:
\begin{itemize}
\item $\qM(D)$ is a von Neumann algebra and the net $D \mapsto \qM(D)$ satisfies conditions mentioned above. Moreover, the vacuum $\Omega$ is cyclic for each $\qM(D)$.
\item Each operator $\phi \in \cP(D)$ has a closed extension $\phi_e \subset \phi^{\dagger, *}$
which is affiliated with $\qM(D)$. Here, $\phi_e \subset A$ means that the domain of $\phi_e$ is contained in the domain of $A$ and that $\phi_e = A$ on the domain of $\phi_e$.
\end{itemize}
\end{theorem}
Theorem \ref{2.3a} yields
\begin{corollary}
Field operators lead to operators affiliated to the von Neumann algebra $\qM(D)$. We remind that this property is the starting point for the definition of measurable operators.
\end{corollary}
But, as it was mentioned in the first section, for large systems, $\qM(D)$ is type III algebra. Therefore, one should employ larger algebra $\cM \equiv \mathfrak{M} \rtimes_{\sigma} \Rn$. It is not too difficult to see ($\mathfrak{M}\subset \cM$) that fields operators lead to operators affiliated to $\cM$. Then, applying the theory of cross-product algebras one gets a nice weight $\tau$ on $\cM$ having the trace property. Consequently, quantum $\tau$-measurable operators $\widetilde{\cM}$ can be defined. In particular, one can define the quantum Orlicz space $L^{cosh - 1}(\qM(D))$. This leads to, see \cite{LM17}:
\begin{definition}
A field operator $\phi(f)$ affiliated to $\mathfrak{M}(D)$ is said to satisfy an \emph{$L^{\cosh-1}$ regularity restriction} if the strong product $\varphi_{\cosh-1}(h)^{1/2}\phi(f)\varphi_{\cosh-1}(h)^{1/2}$ is a closable operator for which the closure is $\tau$-measurable, i.e. the closure is an element of the space $\widetilde{\cM}$, ($h$ is the uniquely determined unbounded operator affiliated to $\cM$ while $\varphi$ stands for the fundamental function associated with the Orlicz space $L^{cosh - 1}$).
\end{definition}

The restriction employed in the above definition is a physically reasonable.  That is to say, membership of $L^{cosh - 1}(\qM(D))$ ensures that the ``generalized moments'' of the field operators are all finite. And this is in perfect agreement with the very essence of the first Wightman postulate. For a fuller treatment we refer the reader to \cite{LM17}.

\subsection{ Quantum correlations.}

The term \textit{quantum correlations} refers to a certain property of quantum states (so linear positive normalized functionals on the algebra specified by the considered system). 
To understand the specificity of that concept, it is natural to begin with  classical probability theory.
In particular, within this theory, the concept of classical correlations is fully described (cf \cite{hal}).
The fundamental ingredients of probability theory are the probability space $(\Omega, \Sigma, \mu)$ and the abelian von Neumann algebra $L^{\infty}(\Omega, \mu)$, where $\Omega$ is a set, $\Sigma$ is a $\sigma$-algebra of measurable subsets of $\Omega$, $\mu$ is a probability measure and $L^{\infty}(\Omega, \mu)$ is the set of all essentially bounded measurable functions. It is important to note here that to get a full description of (classical) correlations, the concepts of a subsystem, a composite system, and product structure of probability spaces are crucial.

The syntax of classical probability theory extends to the noncommutative realm by replacing $L^{\infty}(\Omega, \mu)$ with a von Neumann (in general noncommutative) algebra $\mathfrak{M}$.
We have seen that this was the starting point for noncommutative integration theory. What is more, to quantize the concepts of subsystem and product structure, one is forced to rewrite these concepts in terms of tensor products of appropriate algebras.

However, \textit{when studying correlations, a strong emphasis must be placed on measures}. But, the Riesz-Markov-Kakutani theorem lays down a one-to-one correspondence between (Borel) probability measures on $X$ and a normalized, linear positive functionals (states) on the (abelian $C^*$) algebra $C(X)$ of all complex valued continuous functions on a compact Hausdorff space $X$. Consequently, instead of considering a probability measure one can study the corresponding state.
An analysis of correlations in terms of states vis-a-vis measures has a huge advantage as the description based on states can be directly extended to the noncommutative realm.

Therefore, we must replace $C(X)$ by a $C^*$-algebra $\mathfrak{A}$, with the starting point for a study of quantum correlations being a pair $\left( \mathfrak{A}, \varphi \right)$, where $\varphi$ is a state on $\mathfrak{A}$. In that way, a very general description of ``quantization'' of classical correlations was done, see \cite{Maj1}, \cite{Maj2}, \cite{Maj3} and \cite{Maj4}. In particular, loosely speaking, one can consider the entanglement (frequently considered as a central feature of quantum theory) as a lack of weak-$^*$ Riemann approximation property for a state on products of noncommutative structures. \textit{This clearly indicates the great difference between the standard integration theory and the noncommutative one.}

Frequently, one is interested in states (``quantum measures'') which are somehow related to the stability of  matter in the sense that the number operator is well defined. On the other hand, the existence of the number operator in the corresponding GNS representation characterizes the normality of states, cf Section 5.2.3 in \cite{BR}. However, to speak about normality of a state, the von Neumann algebra setting must be used. In other words, very specific $C^*$-algebras (weakly closed) become the basic ingredient of the description. An important consequence of this algebra replacement,
is that to describe a composite system consisting of two subsystems a very special tensor product - the operator space tensor product - must be employed. This product leads to a proper geometry of density matrices of composite systems. The principal significance of that geometrical characterization is that it allows one to see essential differences between a $C^*$-algebra and von Neumann algebra approach to a characterization of entanglement.

Working within such schemes we obtained, (for details see \cite{Maj4}, \cite{MMO}):

\begin{enumerate}
\item canonical form of two point correlation functions,
\item general definition of entangled states in terms of $C^*$-algebras as well as von Neumann algebras,
\item measures of entanglement for genuine quantum systems,
\item a general description of entanglement of formation,
\item a general characterization of PPT states.
\end{enumerate}

We want to complete this subsection with some remarks on Bell's inequalities.
It should be reminded the reader that frequently quantum correlations are defined by means of these inequalities. As the presented approach is different one, this point needs some clarifications.

In 1964 J. Bell \cite{bell} described an \textit{gedankenexperiment} test for the existence of hidden variables. It deals directly with measurements of variables \textit{obeying classical probability calculus!} In particular, Bell's inequalities concern measurements of particles that have interacted and then separated. These inequalities are given in terms of certain combinations of two point correlation functions. It is worth pointing out that although Bell posited the existence of some hidden variables, there is a proof of Bell's inequalities without any assumption about existence and properties of hidden variables, see \cite{land}, \cite{tsi}.

These inequalities were tested for quantum particles and found to be violated. We must add that such violation was expected for objects governed by quantum mechanics. BUT, such violations mean that correlations of classical objects can be different from the quantum ones! However, the aforesaid setting does not give any explanation about the nature of this phenomenon. One can only say that the classical probability is not enough for the ``quantum world''.

Our approach is giving a clarification. We have shown that quantum correlations are just the result of the quantization procedure. Moreover, our approach offers the natural measures of quantum correlations for genuine quantum systems.

Consequently, Bell's inequalities are important ingredient of the theory of quantum correlations, especially in quantum information, but they cannot be taken as the starting point of the theory. For a deeper discussion on quantization of correlations we refer the reader to \cite{Maj4}.

\section{Final remarks and open problems}
We argued that an applications of the noncommutative calculus for a description of large systems is both natural and fruitful. In particular, the quantization cannot be carried out within finite dimensional structures, and for the case of large systems, does not lead to the one universal algebra $B(\cH)$. Fortunately, the noncommutative theory of integration as well as the theory of derivations are well defined for general von Neumann algebras (even with some extensions for $C^*$-algebra setting).

Needless to say, noncommutative derivatives are also playing an important role in applications of noncommutative calculus to an analysis of quantum systems - for a deeper discussion see \cite{BR}, \cite{sakai}. As an example, how it can trivially be adapted for studies of concrete models, see \cite{MK}, and \cite{MM}.

We close this section with some open problems. The reason is that
quantum statistical mechanics, besides its foundational role in Physics, is currently being enriched with many new problems inspired from an analysis of concrete models. To illustrate this, we wish to list a couple of open problems which (of course) reflects the author's personal taste:
\begin{enumerate}
\item to complete theory of measures of entanglement in terms of von Neumann algebras, cf Remark 8.6 in \cite{Maj4}.
\item to elaborate log-Sobolev inequalities in the generalized framework of quantum statistical mechanics, cf. \cite{RR}, \cite{Gros}, \cite{BG}.
\item using log Sobolev machinery, describe the stability of quantum dynamics and make an analysis of ``return to equilibrium'', cf. \cite{Gros}, \cite{RR}, \cite{Arnold}, \cite{Barthe}, \cite{Kos}.
\item to study Bell's inequalities within the framework of quantum Orlicz spaces; this is a modification of the Tsirelson problem \cite{Tsi1}.
\item Provide simple examples of differential structures for Quantum Field Theory, cf \cite{LM17}.
\end{enumerate}

\section{Acknowledgments}
I am very grateful to L.E. Labuschagne, who helped me a lot in correcting these notes. The author is greatly indebted to the referee for drawing the author's attention to $O^*$-algebras.

\end{document}